\documentclass{emulateapj}
\shorttitle{\textit{Spitzer} Upper Limits on AXPs}
\shortauthors{Wang, Kaspi, \& Higdon}

\newcommand{\chandra}{\textit{Chandra}}
\newcommand{\spitzer}{\textit{Spitzer}}
\newcommand{\eaxp}{1E~1048.1$-$5937}
\newcommand{\rxs}{1RXS~J170849.0$-$400910}
\newcommand{\xte}{XTE~J1810$-$197}

\begin{document}
%\bibliographystyle{apj_noskip}

%\submitted{Submitted to ApJ}
\title{\textit{Spitzer} Mid-infrared Upper Limits on Anomalous
X-Ray Pulsars 1E 1048.1$-$5937, 1RXS J170849.0$-$400910, and XTE J1810$-$197}

\author{Zhongxiang Wang\altaffilmark{1},
Victoria M. Kaspi\altaffilmark{1}, and Sarah J. U. Higdon\altaffilmark{2,3}
} 
\altaffiltext{1}{\footnotesize Department of Physics,
McGill University, Montreal, QC H3A 2T8, Canada;
wangzx@physics.mcgill.ca, vkaspi@physics.mcgill.ca}
\altaffiltext{2}{Department of Astronomy, 
Cornell University, Ithaca, NY 14853}
\altaffiltext{3}{Department of Physics,  
Georgia Southern University, Statesboro, GA 30460;
shigdon@georgiasouthern.edu}
%%\email{wangzx@physics.mcgill.ca, vkaspi@physics.mcgill.ca, }

\begin{abstract}
We report on mid-infrared imaging observations of the anomalous 
X-ray pulsars (AXPs) \eaxp, \rxs, and \xte.  The observations were carried out
at 4.5 and 8.0 $\mu$m with the Infrared Array Camera and
at 24 $\mu$m with the Multiband Imaging Photometer 
on the \textit{Spitzer Space Telescope}.  No mid-infrared counterparts
were detected.  As infrared emission from AXPs may 
be related to their X-ray emission either via the magnetosphere or 
via a dust disk,  we compare the derived upper limits on 
the infrared/X-ray flux ratios of the AXPs to the same ratio for 
4U~0142+61, an AXP previously detected in the mid-infrared range.  
The upper limits are above the flux ratio for
4U 0142+61, indicating that if AXPs have similar infrared/X-ray flux ratios,
our observations were not sufficiently deep to detect our AXP targets. 
For \xte, the upper limits set a constraint on its rising radio 
energy spectrum, suggesting a spectral break between 
4.2$\times 10^{10}$ and 6$\times$10$^{13}$ Hz.  

\end{abstract}


\section{Introduction}

The anomalous X-ray pulsars (AXPs) are a small group of isolated, young
neutron stars with spin periods falling in a narrow range (5--12~s), 
very soft X-ray spectra in the 0.5--10 keV range,  
and with X-ray burst activity (\citealt{wt04}; \citealt{kas06}). 
They were considered ``anomalous'' because their X-ray luminosities greatly
exceed the power available from the rotational kinetic
energy of the pulsars.  Along with
soft gamma-ray repeaters (SGRs),  these
objects are believed to be neutron stars having
extremely strong ($\sim10^{14}$ G) surface magnetic fields
(``magnetars''; \citealt{td96}). 

In addition to studies at X-ray energies,  effort has
been made to observe AXPs at optical and infrared (IR)
wavelengths.  Since the discovery of the optical/near-IR counterpart to 
AXP 4U 0142$+$61 \citep*{hvk00},  four other AXPs, \eaxp\ \citep{wc02},
\rxs\ \citep{dv06a, isr+03}, \xte\ \citep{isr+04}, 
and 1E 2259$+$586 \citep{hul+01},  were also identified in the near-IR. 
The observed near-IR emission,  which greatly exceeds the extrapolated spectrum
of the X-ray blackbody component of AXPs,  is likely related 
to the X-ray emission---AXPs show very similar $K$-band to X-ray flux ratios 
(1.7--3.7 $\times 10^{-4}$; \citealt{dv05, dv06a}) and 
correlated near-IR and X-ray flux variations were found
for 1E~2259+586 \citep{tam+04} and \xte\  \citep{rea+04} during their
X-ray outbursts.

In the magnetar model, optical/IR emission is probably due to
non-thermal radiation by particles in the magnetosphere 
(e.g., \citealt{bt07}).
However, using the \textit{Spitzer Space Telescope}, \citet*{wck06} recently
identified 4U~0142$+$61 at the mid-IR wavelengths 4.5 and 8.0 $\mu$m.  
The emission in the mid-IR was brighter than that in the near-IR, and can be
interpreted as dust emission from an X-ray heated passive disk 
around the young pulsar. Given the similar near-IR fluxes of AXPs,
this suggests that AXPs could all be bright at mid-IR wavelengths, 
due to non-thermal emission from their magnetospheres
or the existence of surrounding disks.

In this paper, we report on the \spitzer\ observations of the AXPs
\eaxp, \rxs, and \xte. The general properties of these three AXPs are summarized
in Table~\ref{tab:axp}.  We present mid-IR observations in
\S~\ref{sec:obs} and our results in \S~\ref{sec:res}.
We discuss the implications of our observations in \S~\ref{sec:disc}.

\section{{\em Spitzer} Observations}
\label{sec:obs}

The fields of three AXPs were observed  with \textit{Spitzer} using the Infrared
Array Camera (IRAC; \citealt{fha+04}) and Multiband Imaging Photometer
for \textit{Spitzer} (MIPS; \citealt{rie+04}), as part of the Infrared
Spectrograph (IRS) instrument team guaranteed time program.  
A summary of the \spitzer\ observations is given in Table~\ref{tab:obs}.

\subsection{\spitzer\  IRAC 4.5 and 8.0 $\mu$m Imaging}

IRAC is an imaging camera operating in four channels at  
3.6, 4.5, 5.8, and 8.0 $\mu$m.  Two adjacent fields
are simultaneously imaged in pairs (3.6 and 5.8 $\mu$m;
4.5 and 8.0 $\mu$m).  We observed our targets in 
the 4.5 and 8.0 $\mu$m channels, with bandwidths of 1.0 and 2.9 $\mu$m,  
respectively. 
The detectors at the short and long wavelengths are InSb
and Si:As devices, respectively, with 256$\times$256 pixels and a plate
scale of 1\farcs2, giving a field of view (FOV) of 5\farcm2$\times$5\farcm2.
The frame time was 100 s, with 96.8 s effective exposure time
per frame for the 4.5 $\mu$m data and 93.6 s effective exposure time
for the 8.0 $\mu$m data. The total exposures of each target's field
are given in Table~\ref{tab:obs}.
%\begin{figure}
\begin{center}
\includegraphics[scale=0.68]{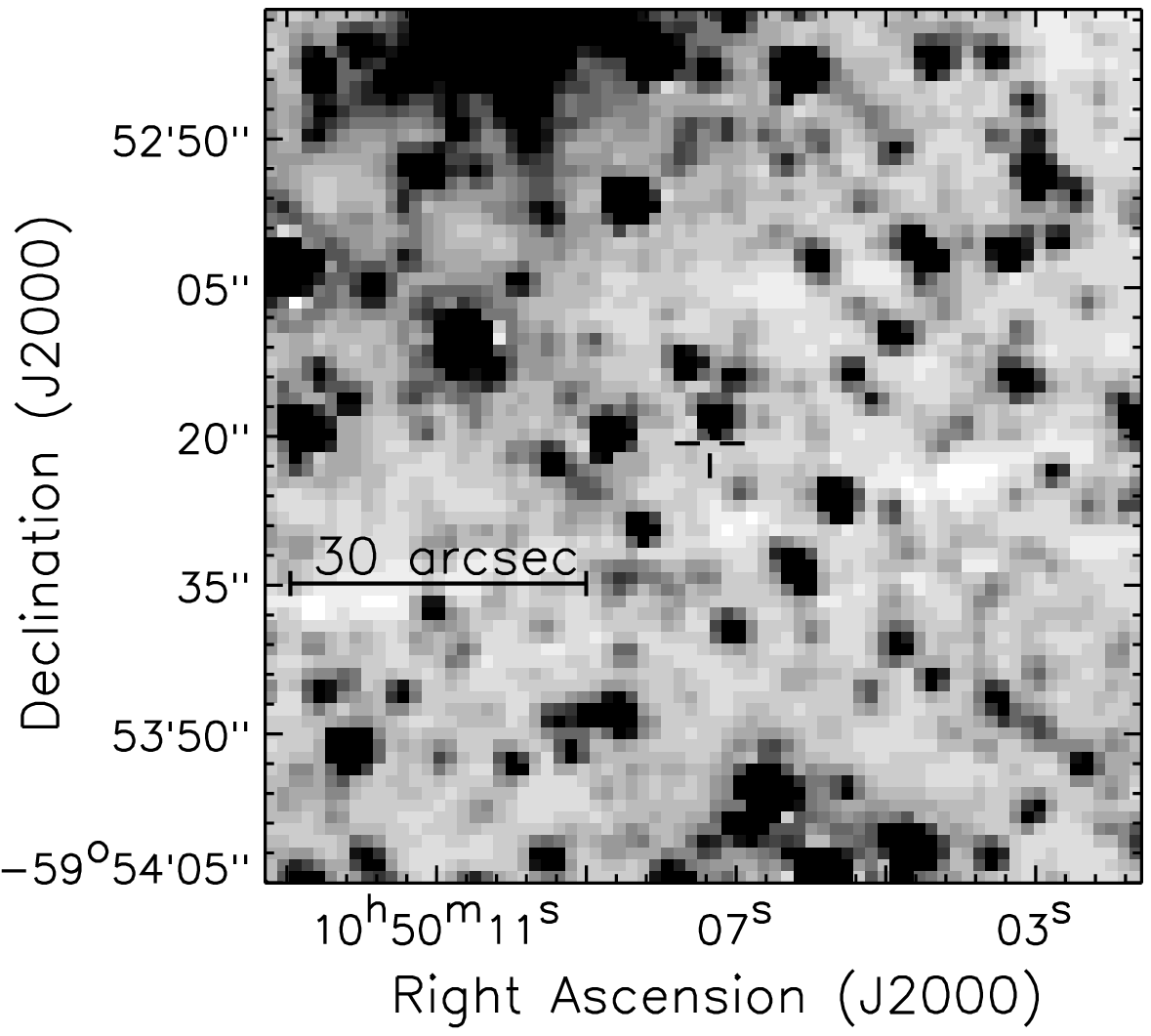}
%\plotone{f1.eps} 
\figcaption{\spitzer\ IRAC 4.5 $\mu$m
image of the \eaxp\  field. The source position \citep{wc02} is indicated by
a cross. No object was found within a 0\farcs5 radius error circle
(too small to be resolved on the images),
which is the uncertainty (90\% confidence) of locating the near-IR source 
position in the \spitzer\  images.
\label{fig:eaxp}}
\end{center}

%\end{figure}
%\begin{figure}
%fig2
\begin{center}
\includegraphics[scale=0.68]{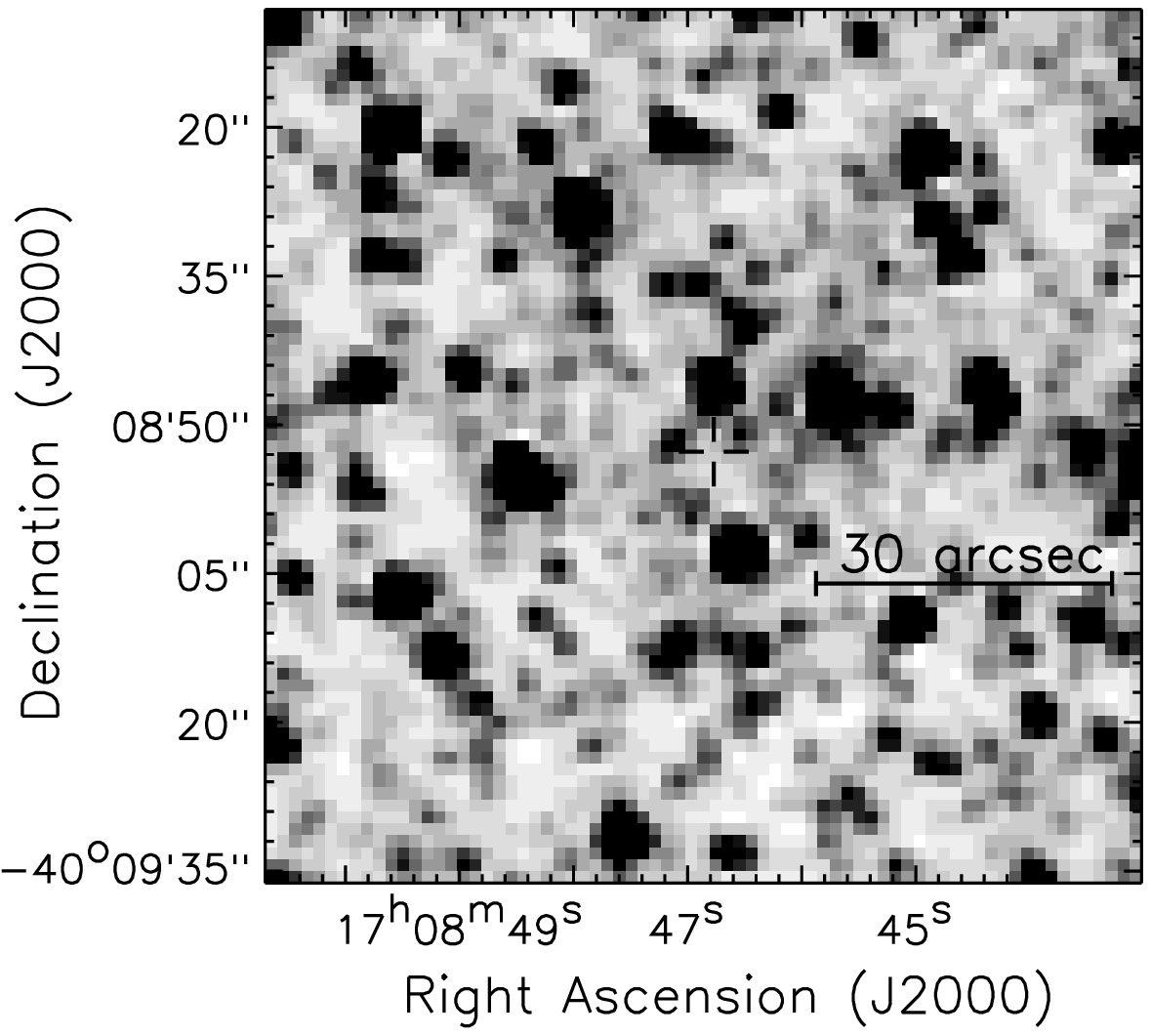}
%\plotone{f2.eps} 
\figcaption{\spitzer\  IRAC 4.5 $\mu$m 
image of the \rxs\  field. The source position
\citep{isr+03}, with a nominal uncertainty of 0\farcs7 (90\% confidence),
is indicated by a cross (the error circle
is too small to be resolved on the images).}
\end{center}
%\end{figure}

The data were processed through the data reduction pipelines (version
S14.0.0; IRAC Data Handbook, version 3.0, 2006) at the {\em Spitzer} Science
Center. In the basic calibrated data (BCD) pipeline, the individual
flux-calibrated frames were produced from the raw images. The BCD
frames were then combined into the post-BCD (PBCD) mosaics. 
The pointing of the IRAC frames is typically accurate to 0\farcs5.
Because the fields of the AXPs are crowded,
we further astrometrically calibrated the PBCD images to
achieve the best positional accuracy and avoid source confusion. 
For the \eaxp, \rxs, and \xte\  images, 
261, 230, and 100 Two Micron All-Sky Survey 
(2MASS; \citealt{2mass}) stars, detected by 
each of the 4.5 $\mu$m images, were used for the calibrations, respectively.
The nominal uncertainties of the calibrated images 
are dominated by the 2MASS systematic uncertainty ($\simeq$0\farcs15, with
respect to the International Celestial Reference System).
%\begin{figure}
%fig3
\begin{center}
\includegraphics[scale=0.68]{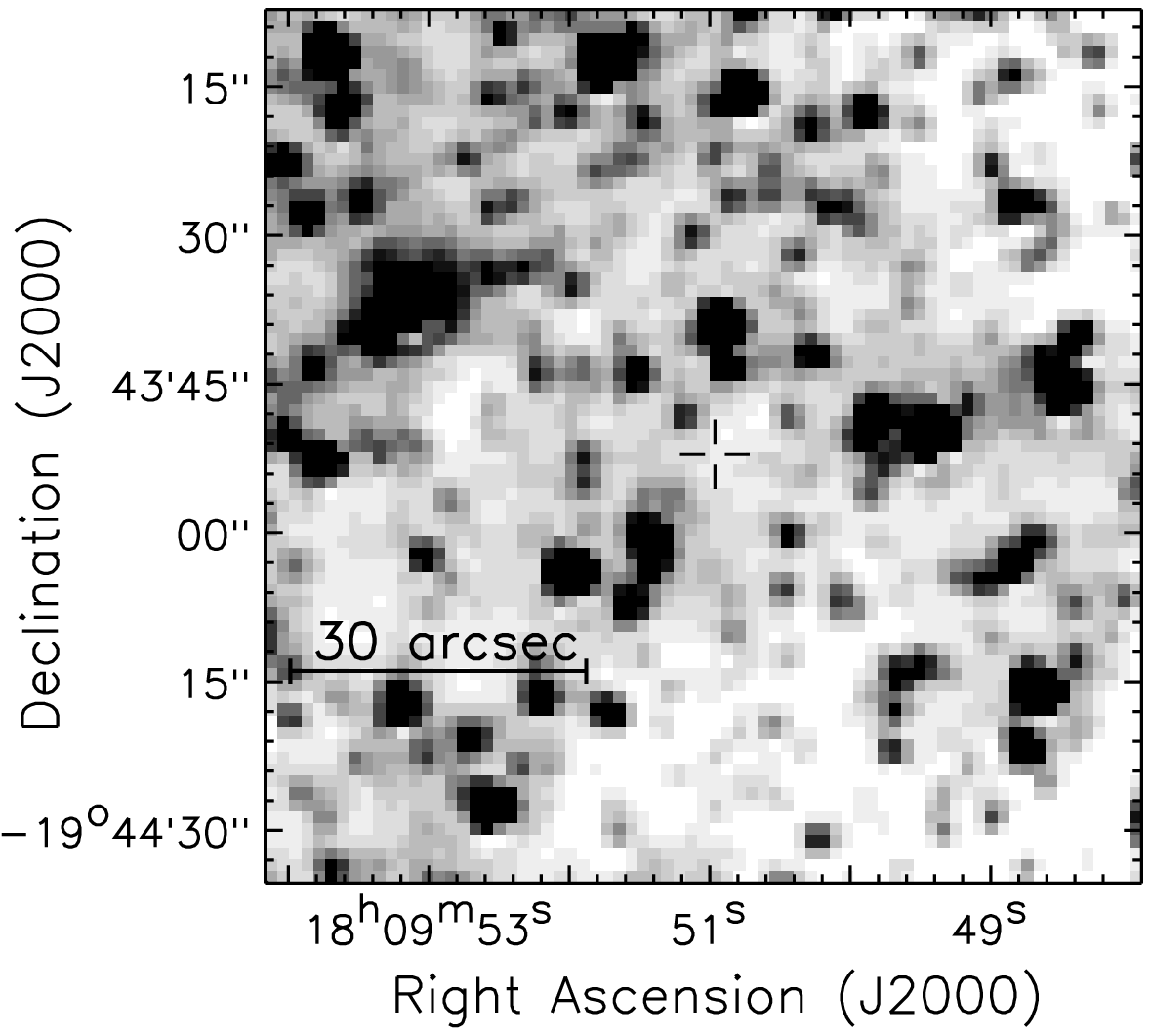}
%\plotone{f3.eps}
\figcaption{\spitzer\  IRAC 4.5 $\mu$m 
image of the \xte\ field. The \chandra\  source position \citep{isr+04}
is indicated by a cross.  The nominal uncertainty 
(90\% confidence) on the source position is 0\farcs7. This error circle 
is too small to be resolved on the images.}
\end{center}
%\end{figure}

\subsection{\spitzer\  MIPS 24 $\mu$m Imaging}

MIPS contains three separate detector arrays which provide
capabilities for imaging and photometry in broad bands at
24, 70, and 160 $\mu$m. The 24 $\mu$m camera (having 4.7 $\mu$m bandwidth)
was operated under the Photometry
and Super Resolution (PH/SR) mode for 
the observations.  The detector is a 128$\times$128 pixel
Si:As array, with a pixel size of 2\farcs55 and a 5\farcm4$\times$5\farcm4
FOV.   The frame time used was 10-s, with 9.96-s effective 
exposure time per frame.  The total exposure of each source, controlled
by the frame time, number of frames per cycle, and number of cycles,
is given in Table~\ref{tab:obs}.

The data were processed through the MIPS data reduction pipelines (version
S13.2.0; MIPS Data Handbook, version 3.2.1, 2006) at the {\em Spitzer} Science
Center. Similarly to the IRAC data processing, 
%and BCD images were produced from the pipelines. From the BCD images,
a PBCD image of each target field was produced by the post-BCD 
pipeline from the BCD images (see MIPS Data Handbook for details).

\section{Results}
\label{sec:res}

In Figures 1--3, we show the mid-IR images  of the \eaxp, \rxs, and \xte\  
fields, respectively.  For the first
AXP, the source position is
from the near-IR observation \citep{wc02}, while for the
latter two, their \chandra\  X-ray positions \citep{isr+03,isr+04} were used. 
Combining the positional uncertainties on
the \spitzer\ images and the sources, the 90\% confidence radii
of the error circles for the AXPs are at 0\farcs5--0\farcs7.
In each figure, the source position is indicated by a cross
(the error circles are too small to be resolved on the images). 
No objects were found within the error circle of each source position 
on the \spitzer\  images.  There is one faint object near the position of \rxs,
but it is 1\farcs8 (or more than 4$\sigma$) away from
the \chandra\  position. 
%\begin{figure}
%%fig4
\begin{center}
\includegraphics[scale=0.83]{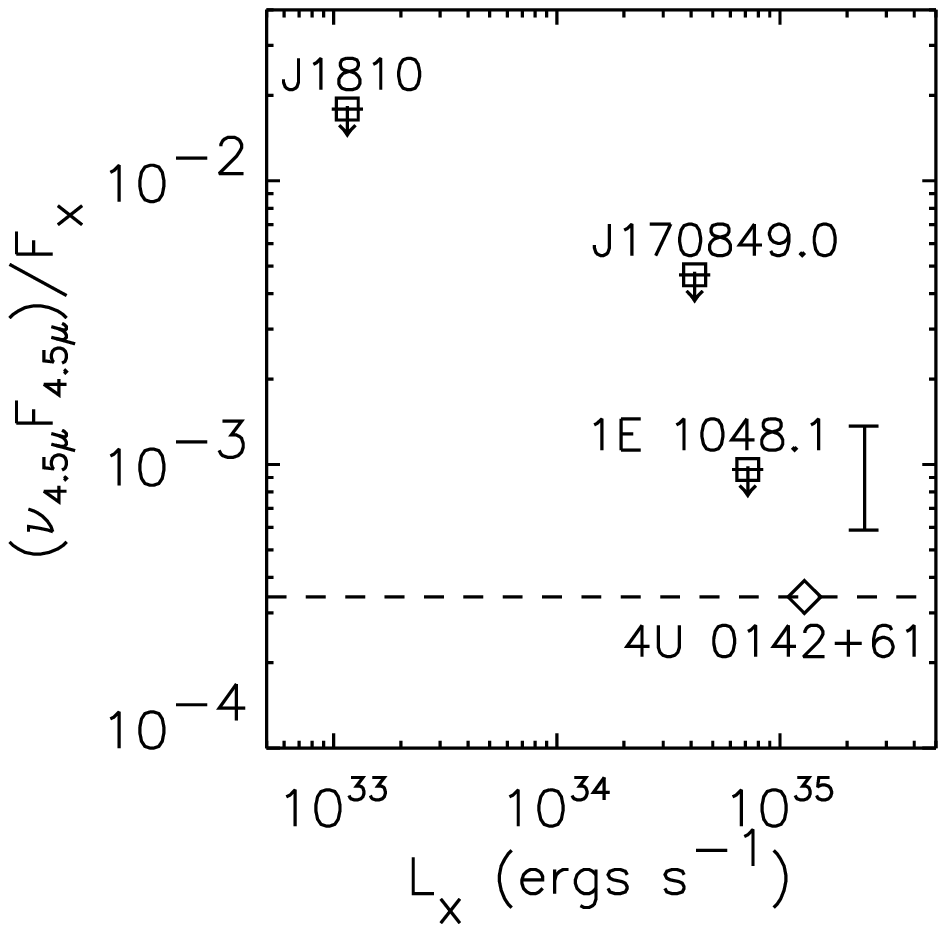}
%\plotone{f4.eps}
\figcaption{Upper limits on the \textit{Spitzer}/IRAC 
4.5~$\mu$m to X-ray (in the 2--10 keV range) flux ratios of the three AXPs. 
Fluxes are unabsorbed, and values of $A_V= 5.6, 7.7$, and 3.6 
(derived from $N_{\rm H}$ by using
$A_V= N_{\rm H}/0.179\times 10^{22}$ cm$^{-2}$;  \citealt{ps95})
are used for dereddening the 4.5~$\mu$m flux upper 
limits \citep{imb+05} of \eaxp, \rxs, and \xte, respectively. 
The dashed line indicates the flux ratio of the AXP 4U~0142+61. 
The error bar near \eaxp\  indicates the uncertainty on its upper
limit due to its X-ray variability \citep{tie+05}.
\label{fig:lim}}
\end{center}
%\end{figure}

The sensitivity of the IRAC observations is dominated by background
sky emission and confusion noise when a field is crowded. The
sky brightnesses measured by IRAC and MIPS
at our target positions are given in Table~\ref{tab:obs}.
As can be seen, \eaxp\  has 
a background between medium to high\footnote{See
www.spitzer.caltech.edu/obs/bg.html} (e.g., at 8 $\mu$m, 
the sky brightness is between 7.7--18.3 MJy/sr).
The backgrounds of the other two AXPs are 
much higher than the high background defined by \spitzer, 
presumably because of their crowded fields.
We derived the 3$\sigma$ flux upper limits from the standard
deviation of the background sky at the source positions. The upper limits
are given in Table~\ref{tab:obs}.

In addition, we note that for the two objects within the error 
circle of \rxs\, which were both proposed as the possible near-IR 
counterpart \citep{isr+03,dv06a}, our IRAC observations would have 
detected the object $A$,  
if we assume that this AXP has the same mid-IR to near-IR flux ratio
as that of 4U 0142+61. 
Therefore, the non-detection of \rxs\  in our IRAC images is also in 
favor of object B being the near-IR counterpart \citep{dv06a}.
%\begin{figure}
%%fig5
\begin{center}
\includegraphics[scale=0.54]{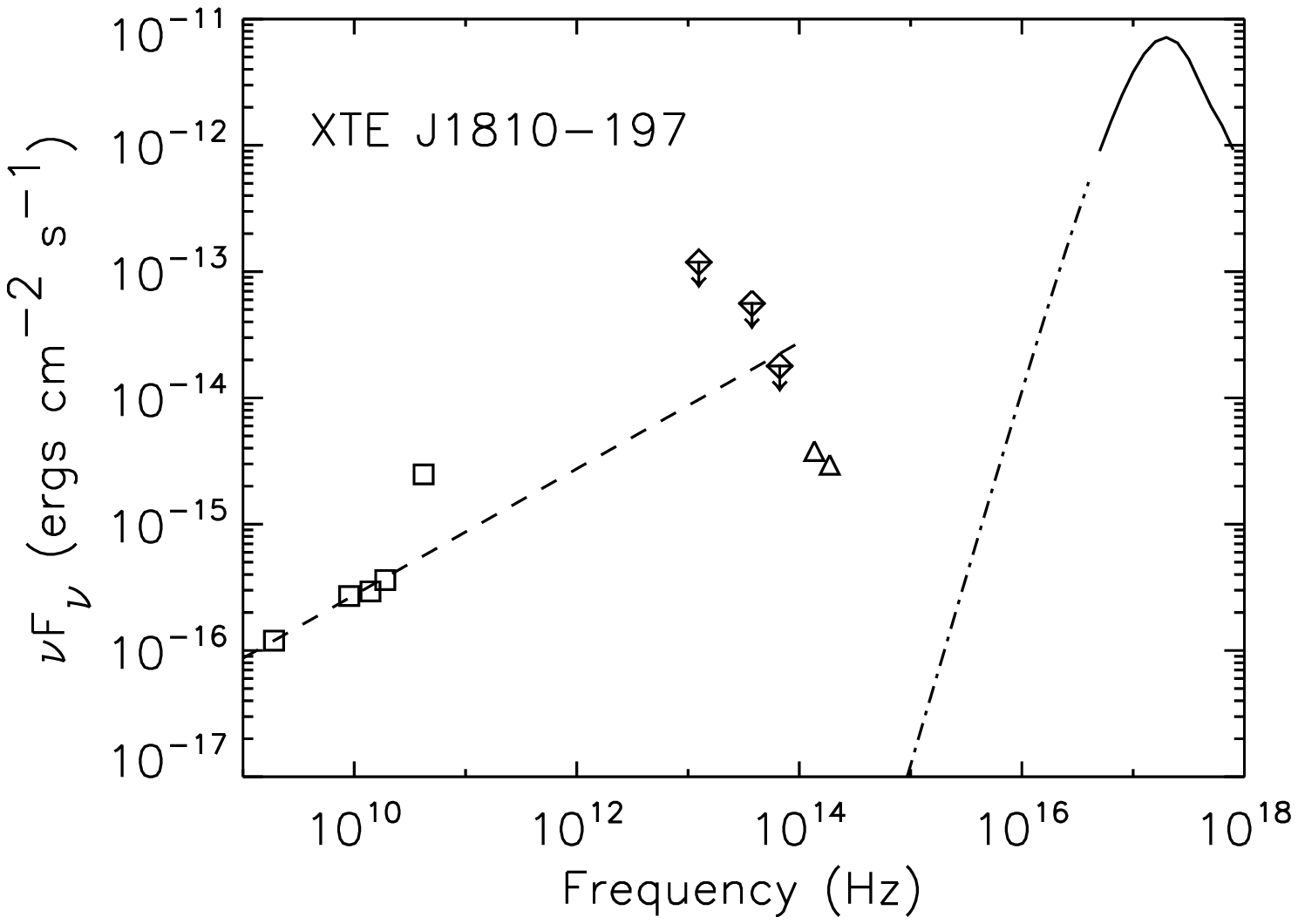}
%%\plotone{f5.eps} 
\figcaption{Unabsorbed mid-IR flux upper limits for \xte\  (diamonds). 
Also shown are its two-blackbody model X-ray flux (solid and dot-dashed curves) 
from the \textit{XMM} observation on 2005 September 20 \citep{gh06}, 
dereddened near-IR $HK_s$ fluxes (triangles) measured on 2004 March
12--14 \citep{rea+04}, and radio 1.9, 9.0, 14, 19 (on 2006 May 2) 
and 42 GHz (on 2006 May 3) fluxes (squares; \citealt{cam+06}). 
An $S_{\nu}\propto \nu^{-0.5}$ radio spectrum, which well represents 
the May-2 data, is indicated by a dashed line.
\label{fig:j18} }
\end{center}
%\end{figure}

\section{Discussion}
\label{sec:disc}

As indicated by the similar $K$-band--to--X-ray flux ratios and 
the cases of correlated near-IR/X-ray flux variations (although
only seen during outbursts), IR emission from AXPs is likely connected
to their X-ray emission, either via the magnetosphere or via reprocessing of
X-rays by a surrounding dust disk. In the first case, 
as starquakes shear the external magnetic field, 
a plasma corona forms in the twisted magnetosphere of a magnetar 
\citep{tlk02, bt07}.  The observed 
blackbody component ($kT\sim 0.5$ keV)
arises from the atmosphere of this
magnetar, which would be partly heated by particles from the corona.
The 2--10 keV power-law component is produced by
multiple cyclotron scattering of the keV tail of the blackbody
by coronal particles.  The observed optical/IR emission is
probably related to the X-rays, as it may be due to cyclotron or
curvature radiation by the particles in the corona \citep{bt07}.
However, currently there is no detailed theoretical model to explain
the similar near-IR/X-ray flux ratios in the magnetosphere picture.
Alternatively, the IR and X-ray emission may be connected
by a dust disk \citep{wck06}. The putative disks would have formed 
from fallback of supernova material \citep*{lwb91,chn00}. 
Similar to an accretion disk in X-ray binaries,
the dust disk is heated by the strong X-rays from the central neutron star,
but since it is farther away from the central source and cooler than 
an accretion disk (e.g., see \citealt{hvk00} for the accretion disk model 
that was previously considered for the optical emission from 4U 0142$+$61),
it emits mainly in the IR \citep*{phn00}. 
Using {\em Spitzer}/IRAC observations similar
to those discussed here, the unabsorbed IR--to--X-ray flux ratio
of 4U 0142+61 was found to have a maximum value of 
$(\nu_{4.5\mu{\rm m}}F_{4.5\mu{\rm m}})/F_{\rm X}
\simeq$3.4$\times 10^{-4}$ at 4.5 $\mu$m (here $F_{\rm X}$ is phase-averaged in 
the 2--10 keV range, $F_{\rm X}=8.3\times 10^{-11}$ ergs cm$^{-2}$ s$^{-1}$
and $N_{\rm H}=0.91\times 10^{22}$ cm$^{-2}$;
\citealt{pat+03}).  This ratio, as it depends
largely on the geometry of the disk and not on the peculiarities of
the central object, might be typical for all dust disks around young
neutron stars.  Thus, as a first-order 
hypothesis for either case, it is not unreasonable to assume that 
the ratio is constant from source to source, and see whether the data 
support this hypothesis.

In Figure~\ref{fig:lim}, we show the upper limits 
on the mid-IR 4.5 $\mu$m to X-ray unabsorbed flux ratios of the three
target AXPs and compare them to the same ratio for 4U 0142+61. 
The unabsorbed X-ray fluxes we use are given in Table~\ref{tab:axp}.
Depending on the models used in spectral fitting, the resulting flux 
in the soft 0.5--2 keV range of 
the AXPs can be different by a factor of 2--3 (e.g., \citealt{tie+05}). 
We therefore use the 2--10 keV unabsorbed flux, which does not  
change significantly for different fitting models, for comparison.
As can be seen from the figure, all the upper limits are 
above the ratio of 4U 0142+61. (This is also true 
when the 0.5--10 keV X-ray fluxes are used for the comparison.)
The non-detection of the sources in the mid-IR, partly because of
the relatively low sensitivity limits of the \spitzer\  observations
(see Table~\ref{tab:obs}), could be due to their relatively 
low X-ray fluxes.

Our mid-IR upper limits on \xte\  set a constraint on
its spectral energy distribution (SED). Its radio spectrum 
was found to be harder ($S_{\nu}\propto \nu^{\alpha}$, $\alpha\gtrsim -0.5$)
than that of most pulsars (\citealt{cam+06}). In Figure~\ref{fig:j18}, 
we show our dereddened upper limits
with the reported fluxes from radio to X-ray energies. We used $A_V\approx 3.6$,
derived from its $N_{\rm H}=0.65\times 10^{22}$ cm$^{-2}$ \citep{gh06}
by using $A_V=N_{\rm H}/(1.79\times 10^{21}$ cm$^{-2})$ \citep{ps95},
to deredden the IRAC data and MIPS datum (according to the reddening laws
of \citealt{imb+05} and \citealt{wd01}, respectively). 
As can be seen, the rising
radio spectrum of the source ($\alpha =-0.5$, estimated from the radio fluxes
measured nearly simultaneously) is constrained by 
the 4.5 $\mu$m upper limit: a spectral break is thus expected to be between 
4.2$\times$10$^{10}$ and 6$\times$$10^{13}$ Hz. In addition, more recently,
a flux density of 1.2 mJy at 144 GHz 
(corresponding to 1.7$\times$10$^{-15}$ ergs cm$^{-2}$ s$^{-1}$) from
the source was reported (Camilo et al. 2007). This measurement
further narrows the spectral break range, and may suggest a harder
spectrum (e.g., $\alpha=-0.3$). However, since
the AXP is highly variable in every spectral window where it has 
been observed and the measurements at the different
frequency ranges in Figure~\ref{fig:j18} were not simultaneously obtained, 
the overall shape of the SED may still be highly uncertain.
It is not clear whether or not the radio
emission is closely related to the observed near-IR 
emission  (see Figure~\ref{fig:j18}).
Even in the magnetar model, the optical/IR emission would arise from 
the corona of the magnetar, which is located at the closed magnetosphere region.
Based on the rising radio energy spectrum, deeper observations at 
$\geq$24~$\mu$m wavelengths, with a sensitivity comparable to 
the upper limit at 4.5 $\mu$m, are required to probe whether there is
radio-related IR emission from this source. A non-detection, on the other
hand, would further constrain the frequency range for the spectral break.

\acknowledgements 
We thank David Kaplan and Wynn Ho for useful discussions.
This research was supported by NSERC via a Discovery Grant, Steacie Supplement,
and by the FQRNT and CIAR.  VMK holds the Lome Trottier Chair, 
a Canada Research Chair and is a R. Howard Webster Foundation Fellow of 
the CIAR.  This work is based [in part] on observations made
with the Spitzer Space Telescope, which is operated by the Jet
Propulsion Laboratory, California Institute of Technology under NASA
contract 1407. Support for this work was provided by NASA through
Contract Number 1257184 issued by JPL/Caltech.
This research has made use of the
data products from the Two Micron All Sky Survey, which is a joint
project of the University of Massachusetts and the Infrared Processing
and Analysis Center/Caltech, funded by NASA and NSF.

\bibliographystyle{apj}
%%\bibliography{axp}

\begin{thebibliography}{41}
\expandafter\ifx\csname natexlab\endcsname\relax\def\natexlab#1{#1}\fi

\bibitem[{{Beloborodov} \& {Thompson}(2007)}]{bt07}
{Beloborodov}, A.~M. \& {Thompson}, C. 2007, ApJ, 657, 967

\bibitem[{{Camilo} {et~al.}(2006){Camilo}, {Ransom}, {Halpern}, {Reynolds},
  {Helfand}, {Zimmerman}, \& {Sarkissian}}]{cam+06}
  {Camilo}, F., {Ransom}, S.~M., {Halpern}, J.~P., {Reynolds}, J., {Helfand},
    D.~J., {Zimmerman}, N., \& {Sarkissian}, J. 2006, \nat, 442, 892

\bibitem[{Camilo} {et~al.}(2007)]{cam+07}
  {Camilo}, F., {et~al.} 2007, Submitted to ApJ, (arXiv:0705.4095)

\bibitem[{Chatterjee {et~al.}(2000)Chatterjee, Hernquist, \& Narayan}]{chn00}
    Chatterjee, P., Hernquist, L., \& Narayan, R. 2000, \apj, 534, 373

\bibitem[{{Durant} \& {van Kerkwijk}(2005)}]{dv05}
    {Durant}, M. \& {van Kerkwijk}, M.~H. 2005, \apj, 627, 376

\bibitem[{{Durant} \& {van Kerkwijk}(2006{\natexlab{a}})}]{dv06a}
    ---. 2006{\natexlab{a}}, \apj, 648, 534

\bibitem[{{Durant} \& {van Kerkwijk}(2006{\natexlab{b}})}]{dv06b}
    ---. 2006{\natexlab{b}}, ApJ, 650, 1070

\bibitem[{{Fazio} {et~al.}(2004)}]{fha+04}
{Fazio}, G.~G. {et~al.} 2004, \apjs, 154, 10

%\bibitem[{{Gavriil} {et~al.}(2006){Gavriil}, {Kaspi}, \& {Woods}}]{gkw06}
%{Gavriil}, F.~P., {Kaspi}, V.~M., \& {Woods}, P.~M. 2006, \apj, 641, 418

\bibitem[{{Gotthelf} \& {Halpern}(2006)}]{gh06}
{Gotthelf}, E.~V. \& {Halpern}, J.~P. 2006, in "Isolated Neutron Stars", 
Astrophysics \& Space Science, 308, 79, (astro-ph/0608473)

%\bibitem[{{Halpern} \& {Gotthelf}(2005)}]{hg05}
%  {Halpern}, J.~P. \& {Gotthelf}, E.~V. 2005, \apj, 618, 874

\bibitem[{{Hulleman} {et~al.}(2001){Hulleman}, {Tennant}, {van Kerkwijk},
    {Kulkarni}, {Kouveliotou}, \& {Patel}}]{hul+01}
    {Hulleman}, F., {Tennant}, A.~F., {van Kerkwijk}, M.~H., {Kulkarni}, S.~R.,
      {Kouveliotou}, C., \& {Patel}, S.~K. 2001, \apjl, 563, L49

\bibitem[{{Hulleman} {et~al.}(2000){Hulleman}, {van Kerkwijk}, \&
      {Kulkarni}}]{hvk00}
{Hulleman}, F., {van Kerkwijk}, M.~H., \& {Kulkarni}, S. 2000, \nat, 408, 689

%\bibitem[{{Hulleman} {et~al.}(2004){Hulleman}, {van Kerkwijk}, \&
  %{Kulkarni}}]{hvk04}
  %{Hulleman}, F., {van Kerkwijk}, M.~H., \& {Kulkarni}, S.~R. 2004, \aap, 416,
    %1037

%\bibitem[{{Ibrahim} {et~al.}(2004)}]{ibr+04}
%    {Ibrahim}, A.~I. {et~al.} 2004, \apjl, 609, L21

\bibitem[{{Indebetouw} {et~al.}(2005)}]{imb+05}
    {Indebetouw}, R. {et~al.} 2005, \apj, 619, 931

%\bibitem[{{Israel} {et~al.}(2002)}]{isr+02}
%    {Israel}, G.~L. {et~al.} 2002, \apjl, 580, L143

\bibitem[{{Israel} {et~al.}(2003)}]{isr+03}
    {Israel}, G.~L. {et~al.} 2003, \apjl, 589, L93

\bibitem[{{Israel} {et~al.}(2004)}]{isr+04}
    ---. 2004, \apjl, 603, L97

\bibitem[{{Kaspi}(2007)}]{kas06}
{Kaspi}, V.~M. 2007, in Isolated Neutron Stars: From the Interior to the
  Surface, ed. S.~Zane, R.~Turolla, \& D.~Page, Astrophysics \& Space Science,
  308, 1, (astro-ph/0610304)

\bibitem[{{Lin} {et~al.}(1991){Lin}, {Woosley}, \& {Bodenheimer}}]{lwb91}
 {Lin}, D.~N.~C., {Woosley}, S.~E., \& {Bodenheimer}, P.~H. 1991, \nat, 353, 827

%\bibitem[{Lyne \& Graham-Smith(2002)}]{lg02}
% Lyne, A.~G. \& Graham-Smith, F. 2002, in Pulsar Astronomy (Cambridge: Cambridge
%      University Press)

\bibitem[{{Patel} {et~al.}(2003)}]{pat+03}
    {Patel}, S. K. {et~al.} 2003, \apj, 587, 367

%\bibitem[{{Pavlov} {et~al.}(2002){Pavlov}, {Zavlin}, \& {Sanwal}}]{pzs02}
%   {Pavlov}, G.~G., {Zavlin}, V.~E., \& {Sanwal}, D. 2002, in Proceedings of the
%   270.\ WE-Heraeus Seminar on Neutron Stars, Pulsars, and Supernova Remnants,
%  ed. W.~Becker, H.~Lesch, \& J.~Tr\"{u}mper (Garching: MPE), 273,
	    %(astro-ph/0206024)

\bibitem[{{Perna} {et~al.}(2000){Perna}, {Hernquist}, \& {Narayan}}]{phn00}
	    {Perna}, R., {Hernquist}, L., \& {Narayan}, R. 2000, \apj, 541, 344

\bibitem[{{Predehl} \& {Schmitt}(1995)}]{ps95}
	    {Predehl}, P. \& {Schmitt}, J.~H.~M.~M. 1995, \aap, 293, 889

\bibitem[{{Rea} {et~al.}(2005){Rea}, {Oosterbroek}, {Zane}, {Turolla},
  {M{\'e}ndez}, {Israel}, {Stella}, \& {Haberl}}]{rea+05}
  {Rea}, N., {Oosterbroek}, T., {Zane}, S., {Turolla}, R., {M{\'e}ndez}, M.,
    {Israel}, G.~L., {Stella}, L., \& {Haberl}, F. 2005, \mnras, 361, 710

\bibitem[{{Rea} {et~al.}(2004)}]{rea+04}
    {Rea}, N. {et~al.} 2004, \aap, 425, L5

\bibitem[{{Rieke} {et~al.}(2004)}]{rie+04}
    {Rieke}, G.~H. {et~al.} 2004, \apjs, 154, 25

%\bibitem[{{Seward} {et~al.}(1986){Seward}, {Charles}, \& {Smale}}]{scs86}
%    {Seward}, F.~D., {Charles}, P.~A., \& {Smale}, A.~P. 1986, \apj, 305, 814

\bibitem[{{Skrutskie} {et~al.}(2006)}]{2mass}
    {Skrutskie}, M.~F. {et~al.} 2006, \aj, 131, 1163

%\bibitem[{{Sugizaki} {et~al.}(1997){Sugizaki}, {Nagase}, {Torii}, {Kinugasa},
%{Asanuma}, {Matsuzaki}, {Koyama}, \& {Yamauchi}}]{sug+97}
%      {Sugizaki}, M., {Nagase}, F., {Torii}, K., {Kinugasa}, K., {Asanuma}, T.,
%        {Matsuzaki}, K., {Koyama}, K., \& {Yamauchi}, S. 1997, \pasj, 49, L25

\bibitem[{{Tam} {et~al.}(2004){Tam}, {Kaspi}, {van Kerkwijk}, \&
  {Durant}}]{tam+04}
  {Tam}, C.~R., {Kaspi}, V.~M., {van Kerkwijk}, M.~H., \& {Durant}, M. 2004,
    \apjl, 617, L53

%\bibitem[{{Temim} {et~al.}(2006)}]{tem+06}
%    {Temim}, T. {et~al.} 2006, \aj, 132, 1610

\bibitem[{{Thompson} \& {Duncan}(1996)}]{td96}
    {Thompson}, C. \& {Duncan}, R.~C. 1996, \apj, 473, 322

\bibitem[{{Thompson} {et~al.}(2002){Thompson}, {Lyutikov}, \&
      {Kulkarni}}]{tlk02}
      {Thompson}, C., {Lyutikov}, M., \& {Kulkarni}, S.~R. 2002, \apj, 574, 332

\bibitem[{{Tiengo} {et~al.}(2005){Tiengo}, {Mereghetti}, {Turolla}, {Zane},
        {Rea}, {Stella}, \& {Israel}}]{tie+05}
{Tiengo}, A., {Mereghetti}, S., {Turolla}, R., {Zane}, S., {Rea}, N., {Stella},
	  L., \& {Israel}, G.~L. 2005, \aap, 437, 997

%\bibitem[{{Voges} {et~al.}(1998)}]{vog+98}
%{Voges}, W. {et~al.} 1998, in IAU Symp. 179: New Horizons from Multi-Wavelength
%  Sky Surveys, ed. B.~J. {McLean}, D.~A. {Golombek}, J.~J.~E. {Hayes}, \& H.~E.
%    {Payne}, 433

\bibitem[{{Wang} \& {Chakrabarty}(2002)}]{wc02}
    {Wang}, Z. \& {Chakrabarty}, D. 2002, \apjl, 579, L33

\bibitem[{{Wang} {et~al.}(2006){Wang}, {Chakrabarty}, \& {Kaplan}}]{wck06}
    {Wang}, Z., {Chakrabarty}, D., \& {Kaplan}, D.~L. 2006, \nat, 440, 772

\bibitem[{{Weingartner} \& {Draine}(2001)}]{wd01}
    {Weingartner}, J.~C. \& {Draine}, B.~T. 2001, \apj, 548, 296

\bibitem[{Woods \& Thompson(2006)}]{wt04}
  Woods, P.~M. \& Thompson, C. 2006, in Compact Stellar X-ray Sources, ed.
    W.~H.~G. Lewin \& M.~{van der Klis} (Cambridge: Cambridge Univ. Press),
    547, (astro-ph/0406133)

%\bibitem[{{Zavlin} \& {Pavlov}(2004)}]{zp04}
%	{Zavlin}, V.~E. \& {Pavlov}, G.~G. 2004, \apj, 616, 452

\end{thebibliography}

\clearpage
\begin{deluxetable}{lcccccccc}
\tabletypesize{\scriptsize}
\tablewidth{0pt}
\tablecaption{Properties of \eaxp, \rxs, and \xte\label{tab:axp}}
\tablehead{
\colhead{Source} & \colhead{$P$} &  \colhead{$d$}
& \colhead{$N_{\rm H}$/$10^{22}$} & \colhead{$F_{\rm X}$\tablenotemark{a} /$10^{-11}$} & 
\colhead{Adopted Position\tablenotemark{b}} & \colhead{Refs}\\
\colhead{} & \colhead{(s)}  & \colhead{(kpc)}  & \colhead{(cm$^{-2}$)} & \colhead{(ergs cm$^{-2}$ s$^{-1}$)} & &\colhead{} }
\startdata
\eaxp & 6.4 & 8.6 & 1.0 & 0.74 &
10$^{\rm h}$50$^{\rm m}$07\fs13 $-$59\arcdeg53\arcmin21\farcs3
& 1--3  \\
\rxs & 11.0 & 3.8 & 1.38 & 2.4 &
17$^{\rm h}$08$^{\rm m}46\fs90$ $-$40\arcdeg08\arcmin52\farcs64 &
2,4,5 \\
\xte & 5.5 & 3.3 & 0.65 & 0.10 &
18$^{\rm h}$09$^{\rm m}51\fs08$ $-$19\arcdeg43\arcmin51\farcs74 & 6--8 \\
\enddata
\tablerefs{
(1) \citet{tie+05};
(2) \citet{dv06b};
(3) \citet{wc02};
(4) \citet{rea+05};
(5) \citet{isr+03};
(6) \citet{gh06};
(7) \citet{isr+04};
(8) \citet{cam+06}.}

\tablenotetext{a}{Unabsorbed, phase-averaged 2--10 keV X-ray flux.}

\tablenotetext{b}{Source positions are all J2000.0, and have a 90\%
uncertainty of 0\farcs5 for \eaxp\ and 0\farcs7 for \rxs\  and 
\xte.}

\tablecomments{X-ray observations made at epochs close to
the epochs of our \spitzer /IRAC observations (see Table~\ref{tab:obs})
are used.  The X-ray fluxes of \eaxp, \rxs, and \xte\  were 
measured on 2004 July 08 \citep{tie+05}, 2003 Aug 28  
\citep{rea+05}, and 2005 Sep 20 \citep{gh06}, respectively.}
\end{deluxetable}

\clearpage
\begin{deluxetable}{l c c c c c c c c c c c}
\tablewidth{0pt}
\tabletypesize{\footnotesize}
\tablecaption{\spitzer\ Imaging Observations of Three AXPs at 4.5, 8.0, and 24 $\mu$m\label{tab:obs}}
\tablehead{
\colhead{Object} & \colhead{Date}  &  \colhead{Instrument}  &
\multicolumn{3}{c}{Exposure} & \multicolumn{3}{c}{Sky Brightness} &
\multicolumn{3}{c}{Flux upper limit} \\
		 &		   &			    &
\multicolumn{3}{c}{(min)}    & \multicolumn{3}{c}{(MJy/sr)}       &
\multicolumn{3}{c}{(mJy)} \\
		 &		   &			&
\colhead{4.5} & \colhead{8.0} &  \colhead{24} &
\colhead{4.5} & \colhead{8.0} &  \colhead{24} &
\colhead{4.5} & \colhead{8.0} &  \colhead{24} \\
}
\startdata
1E1048\tablenotemark{a} & 2005-06-13 & IRAC & 16.3 & 15.8 & \nodata & 0.52 & 10 & \nodata 
						   & 0.008 & 0.041 & \nodata\\
       & 2005-04-02 & MIPS & \nodata & \nodata & 7.2 & \nodata & \nodata & 37 
       						   & \nodata & \nodata & 0.39 \\
J1708\tablenotemark{a}  & 2005-08-24 & IRAC & 16.3 & 15.8 & \nodata & 3.1  & 63 & \nodata 
						   & 0.12 & 0.17 & \nodata\\
       & 2005-04-12 & MIPS & \nodata & \nodata & 7.2 & \nodata & \nodata & 62
       						   & \nodata & \nodata & 0.59\\
J1810\tablenotemark{a}  & 2005-09-22 & IRAC & 32.4 & 31.4 & \nodata & 2.4  & 77 & \nodata
					           & 0.023 & 0.13 & \nodata\\
       & 2005-04-13 & MIPS & \nodata & \nodata & 7.2 & \nodata & \nodata & 86
       						   & \nodata & \nodata & 0.88\\
\enddata
\tablenotetext{a}{Full names are \eaxp, \rxs, and \xte.}
\end{deluxetable}

%%\clearpage
%%\begin{deluxetable}{l c c c c c c}
%%\tablewidth{0pt}
%%\tabletypesize{\footnotesize}
%%\tablecaption{Sky brightnesses and flux upper limits (3$\sigma$) of target AXPs \label{tab:sky}}
%%\tablehead{
%%\colhead{Object} & \multicolumn{3}{c}{Sky Brightness} & \multicolumn{3}{c}{Flux upper limits} \\
%%		 & \multicolumn{3}{c}{(MJy/sr)} & \multicolumn{3}{c}{(mJy)} \\
%%	 & \colhead{4.5 $\mu$m} & \colhead{8.0 $\mu$m} & \colhead{24 $\mu$m} 
%%	 & \colhead{4.5 $\mu$m} & \colhead{8.0 $\mu$m} & \colhead{24 $\mu$m}\\}

%%\startdata
%%\eaxp & 0.52  & 10 & 37 & 0.008 &  0.041 & 0.39 \\
%%\rxs  & 3.1   & 63 & 62 & 0.12  & 0.17  & 0.59 \\
%%\xte  & 2.4   & 77 & 86 & 0.023  & 0.13  & 0.88 \\
%%\enddata
%%\end{deluxetable}

\end{document}